\documentclass[%
 reprint,
superscriptaddress,
 amsmath,amssymb,
 aps,prl,
]{revtex4-2}

\usepackage{graphicx}% Include figure files
\usepackage{dcolumn}% Align table columns on decimal point
\usepackage{bm}% bold math
\usepackage{hyperref}%

\newcommand{\zh}{\bm}

\newcommand{\dee}{{\varepsilon}}

\newcommand{\mx}{m_{e}}

\newcommand{\zhr}{{\zh r}}

\newcommand{\zhp}{{\zh p}}

\newcommand{\zhalpha}{{\zh\alpha}}

\newcommand{\Br}[1]{(\ref{#1})}
\newcommand{\Eq}[1]{Eq.\ (\ref{#1})}

\newcommand{\txt}[1]{{\rm #1}}

\begin{document}

%\preprint{APS/123-QED}

\title{Electron-positron pair production in bound-bound muon transitions}% Force line breaks with \\

\author{Oleg Yu. Andreev}
%\email{o.y.andreev@spbu.ru}
\affiliation{ Institute of Modern Physics, Chinese Academy of Sciences, Lanzhou, China }
%\affiliation{Institute of Modern Physics, Chinese Academy of Sciences, Nanchang road 509, Lanzhou, 730000, Gansu, China}
\affiliation{St. Petersburg State University, 7/9 Universitetskaya nab., St. Petersburg, 199034, St. Petersburg, Russia}
\affiliation{Petersburg Nuclear Physics Institute named by B.P. Konstantinov of National Research Centre ''Kurchatov Institute'', Gatchina, 188300, Leningrad District, Russia}
\author{Deyang Yu}
%\email{d.yu@impcas.ac.cn}
\email{Contact author: d.yu@impcas.ac.cn}
\affiliation{ Institute of Modern Physics, Chinese Academy of Sciences, Lanzhou, China }
\affiliation{ University of Chinese Academy of Sciences, Beijing, China }
%\affiliation{Institute of Modern Physics, Chinese Academy of Sciences, Nanchang road 509, Lanzhou, 730000, Gansu, China}
%\affiliation{University of Chinese Academy of Sciences, Beijing, 100049, Beijing, China}
%
\author{Konstantin N. Lyashchenko}
%\email{}
\affiliation{ Institute of Modern Physics, Chinese Academy of Sciences, Lanzhou, China }
%\affiliation{Institute of Modern Physics, Chinese Academy of Sciences, Nanchang road 509, Lanzhou, 730000, Gansu, China}
\affiliation{Petersburg Nuclear Physics Institute named by B.P. Konstantinov of National Research Centre ''Kurchatov Institute'', Gatchina, 188300, Leningrad District, Russia}
\affiliation{St. Petersburg State University, 7/9 Universitetskaya nab., St. Petersburg, 199034, St. Petersburg, Russia}
\author{Daria M. Vasileva}
%\email{}
\affiliation{Petersburg Nuclear Physics Institute named by B.P. Konstantinov of National Research Centre ''Kurchatov Institute'', Gatchina, 188300, Leningrad District, Russia}
\affiliation{St. Petersburg State University, 7/9 Universitetskaya nab., St. Petersburg, 199034, St. Petersburg, Russia}


%\affiliation{
%\affiliation{
%\affiliation{%


\date{\today}% It is always \today, today,

\begin{abstract}
We explored a distinct mechanism for matter creation via electron-positron pair production during bound-bound transitions in the deexcitation of one-muon ions. For ions with nuclear charges $Z\geq24$, transitions from low-lying excited states to the 1s-muon state can lead to the production of electron-positron pairs. 
We show that the Breit interaction determines the transition probabilities for states with nonzero orbital momentum. Although the $2s$ state is metastable, pair production arises mainly from the decay of the $2p$ states. Thus, the Breit interaction governs electron-positron pair production in bound-bound muon transitions. This process offers a unique opportunity to explore quantum electrodynamics in strong fields, as well as a class of nonradiative transitions involving electron-positron pair production.
\end{abstract}

\maketitle

%\tableofcontents

%\section{\label{sec:level1}First-level heading:\protect\\ The line
%break was forced \lowercase{via} \textbackslash\textbackslash}

%\footnote{Automatically placing footnotes into the bibliography requires using BibTeX to compile the bibliography.};

%%%%%%%%%%%%%%%%%%%%%%%%%%%%%%%%%%%%%%%%%%%%%%%%%%%%%%%%%%%%%%%%%

Muonic ions and atoms represent a significant area of research in modern physics \cite{nagamine2003,GORRINGE201573,Bao_2023}. While muonic ions share many similarities with electronic ions, including a theoretical description that closely parallels electronic systems \cite{mohr1998pr293-227,Oreshkina2022PhysRevResearch.4.L042040}, they exhibit important qualitative differences. These differences arise from the finite lifetime of the muon, nuclear effects such as finite nuclear size and recoil, and the distinct energy scales of muonic ions. The muon, being about 207 times more massive than the electron \cite{RevModPhys.93.025010}, has an orbital radius 100 to 200 times smaller than that of an electron \cite{wheeler1949RevModPhys.21.133,Knyazeva2022PhysRevA.106.012809}. This results in a much stronger interaction with the nucleus, leading to greater energy release during the decay of excited states compared to conventional electronic systems. In particular, the transition energies of muons can be large enough to produce electron-positron pairs, introducing a unique decay channel for matter creation that is not present in bound-bound transitions of few-electron ions.

The decay channels for bound muons were analyzed by Wheeler \cite{wheeler1949RevModPhys.21.133}.  Some of these decay processes are shared with electronic ions, while others are specific to muonic ions. In medium and heavy ions, the primary decay channels for excited states in one-muon ions are radiative decays \cite{wheeler1949RevModPhys.21.133,Knyazeva2022PhysRevA.106.012809}. In light muonic ions, where electrons are present in addition to the muon, Auger decay can play a significant role. In this paper, we investigate an additional decay channel for muonic ions, arising from their specific energy range: pair-production transitions in which the excess energy is used to produce an electron-positron pair. This type of decay for the $2s$ muon state was previously considered in \cite{wheeler1949RevModPhys.21.133}. Using QED framework, we examine the decays of muonic states with principal quantum numbers $n=2$ and $3$, demonstrating that, despite the metastability of the $2s$ state, the dominant transitions for electron-positron pair production are the $2p\to1s$ transitions. The significance of these transitions is largely due to the substantial contribution of the Breit interaction for the $2p$ states.

%{\bf
Testing QED in strong fields is a pressing and widely discussed issue \cite{Morgner2023,Loetzsch2024}. Most of these studies focus on photon emission processes, specifically measurements of photon frequencies that determine energy levels and g-factors.
Investigating transitions where excess energy is carried away by electron-positron pairs is both highly important and promising. Such studies provide a unique opportunity to explore a largely unexplored branch of QED, potentially offering new insights into fundamental processes in strong-field regimes.
%}

Electron-positron pair production in these muonic transitions has not yet been studied experimentally. Previous research into pair production in atomic physics has primarily focused on its occurrence in relativistic collisions of heavy nuclei with atoms (dynamic pair production) \cite{Belkacem1993PhysRevLett.71.1514,Belkacem1994PhysRevLett.73.2432,Belkacem1997PhysRevA.56.2806,Belkacem1998PhysRevA.58.1253,Bethe1954PhysRev.93.768,decker1991PhysRevA.44.2883,decker1992PhysRevA.45.3343}.

The nonradiative electron-positron pair production transitions, can be schematically described as
\begin{eqnarray}
\mu_{i}
&\to&\label{eqn230930n03}
\mu_{f} + e^{-}(\dee)  + e^{+}(\dee')
%\mu_{f} + e^{-}(\dee,\zhp,\mu)  + e^{+}(\dee',\zhp',\mu')
\,.
\end{eqnarray}
The initial ($i$) and final ($f=1s$) states of the muon are denoted as $\mu_i$ and $\mu_{f}$, respectively. The produced electron and positron are described by their energy ($\dee$, $\dee'$),
momentum ($\zhp$, $\zhp'$) and polarization ($\mu$ and $\mu'$) \cite{akhiezer65b}. The energy release during transitions between the different states can be estimated by the (nonrelativistic) Bohr formula for the energy levels of one-electron ions.
This shows that if the atomic number $Z$ is larger or equal to $22$ (the more precise calculation predicts $Z\ge24$), then the energy release becomes greater than $2\mx c^2$, which is enough to produce an electron-positron pair. The energy conservation law reads
\begin{eqnarray}
\Delta\dee^{(\mu)}_{i\to 1s}
\,=\,
\dee^{(\mu)}_{i}-\dee^{(\mu)}_{1s}
&=&\label{eqn321117n01}
\dee  + \dee'
\,>\,2\mx c^2
\,.
\end{eqnarray}

Our study focuses exclusively on electron-positron pair production  where the electron belongs to the continuum part of the Dirac spectrum. First, in an experimental setup, a muon beam collides with atoms, and the muon can be treated as bound to the nucleus without significantly affecting the atomic electrons, since its orbital radius is much smaller. Thus, low-lying electron states are already occupied, and capture into these states is forbidden by the Pauli principle. Second, the contribution of unoccupied bound electron states to the total muon transition probability is relatively small.

We consider the muon transition involving the electron-positron pair production, given by \Br{eqn230930n03}, within the framework of QED theory. The Furry picture is used \cite{furry51}, in which the interaction between particles and the electric field of the atomic nucleus is fully taken into account. In the lowest order of the QED perturbation theory, the amplitude of this process can be written as (in the Feynman gauge)
\cite{akhiezer65b,andreev08pr}
\begin{eqnarray}
A_{i\to f}
&=&\nonumber
e^2\int d\zhr_1 d\zhr_2
\psi^{(-)+}_{\dee,\zhp\mu}(\zhr_1)
\phi^{+}_{f}(\zhr_2)
(1-\zhalpha^{(1)}\zhalpha^{(2)})
\\
&&\label{eqn231117n02}
\times
\frac{e^{i\frac{\omega r_{12}}{c}}}{r_{12}}
\psi^{(-)}_{-\dee',-\zhp'\mu'}(\zhr_1)
\phi^{}_{i}(\zhr_2)
\,,
\end{eqnarray}
where $\omega=(\dee+\dee')/\hbar$, $r_{12}=|\zhr_1-\zhr_2|$, $\zhalpha^{(i)}$ are the Dirac $\alpha$-matrices acting on the wave functions with the argument $\zhr_i$. The wave functions $\phi^{}_{i,f}(\zhr_2)$ represent the muon in its initial and final states. The wave function $\psi^{(-)}_{\dee,\zhp\mu}(\zhr_1)$ describes the emitted electron, and $\psi^{(-)}_{-\dee',-\zhp'\mu'}(\zhr_1)$ corresponds to the emitted positron \cite{akhiezer65b}. The Coulomb interaction is obtained by setting $\omega=0$ and neglecting the $\zhalpha^{(1)}\zhalpha^{(2)}$ term.

The transition probably can be written as
\cite{akhiezer65b}
\begin{eqnarray}
dw^{(e^+e^-)}_{i\to1s}
&=&\nonumber
\frac{2\pi}{\hbar}\delta(\dee^{(\mu)}_{i}-\dee^{(\mu)}_{f}-\dee-\dee')
\\
&&
\times
|A_{i\to f}|^2
\frac{d\zhp\, d\zhp'}{(2\pi\hbar)^6}
\,.
\end{eqnarray}
By integrating over the electron and positron momenta, averaging over the projections of the total angular momentum of the initial muon state, and summing over the final states (including different polarizations of the produced electron and positron, as well as the projections of the muon total angular momentum), we obtain the total transition probability ($w^{(e^+e^-)}$).

We accounted for the nuclear size corrections using the Fermi distribution for the nuclear charge density.
We also included the lowest order of the nuclear recoil correction and the electron vacuum polarization correction within the Uehling approximation \cite{uehling35PhysRev.48.55}.

\begin{table*}[!htb]
 \caption{\label{tab1}
Transition probabilities (in s$^{-1}$) for one-muon ions. The digits in square brackets indicate powers of $10$. The first row lists the initial (excited) states ($i$). The total energies of the $1s$-muon states  ($\dee_{1s}^{(\mu)}$), including the rest mass of muon, are given explicitly for each $Z$ value. The rows labeled $\Delta\dee^{(\mu)}_{i\to1s}$ show the energy differences $\Delta\dee^{(\mu)}_{i\to1s}=\dee^{(\mu)}_{i}-\dee^{(\mu)}_{1s}$ (in $\mx c^2$), where $\dee^{(\mu)}_{i}$ represent the total energies of the initial ($i$) muon states. The rows $W_i$ represent the total radiative transition probability for transitions from state $i$ to all lower states. The rows $w^{(e^+e^-)}_{i\to1s}$ provide the transition probabilities for the nonradiative electron-positron pair production transitions (from excited state $i$ to the $1s$ muon state) as described in \Eq{eqn230930n03}, while $w^{(e^+e^-)\txt{C}}_{i\to1s}$ gives these transition probabilities considering only the Coulomb interaction. The rows $w^{(\gamma,e^+e^-)}_{i\to n\to 1s}$ present the transition probabilities for radiative cascade transitions $i\to n\to 1s$ involving electron-positron pair production as described in \Eq{eqn240917n01}, where $i$ refers to the states listed in the first row.
} 
\begin{ruledtabular}
\begin{tabular}{c|c|c|c|c|c|c|c|c}
&$2p_{1/2}$ & $2p_{3/2}$ & $2s_{1/2}$ & $3p_{1/2}$ & $3d_{3/2}$ & $3p_{3/2}$ & $3d_{5/2}$ & $3s_{1/2}$
\\
\hline
%\\
\multicolumn{9}{c}{$Z=24\,,\qquad \dee_{1s}^{(\mu)}=203.8292$}
\\
$\Delta\dee^{(\mu)}_{i\to1s}$&$2.1362$&$2.1421$&$2.1717$&$2.5833$&$2.5852$&$2.5851$&$2.5858$&$2.5939$
\\
$W_i$& $3.97[16]$&$3.94[16]$&$4.95[12]$&$1.17[16]$&$4.54[15]$&$1.17[16]$&$4.49[15]$&$2.78[14]$  
\\
$w^{(e^+e^-)}_{i\to1s}$& $2.14[11]$&$2.42[11]$&$6.23[7]$&$1.19[12]$&$6.73[8]$&$1.22[12]$&$6.77[8]$&$5.71[8]$  
\\
$w^{(e^+e^-) \txt{C}}_{1s}$&$7.65[10]$&$8.55[10]$&$6.22[7] $&$3.35[11] $&$ 2.21[8]$&$3.45[11] $&$2.23[8] $&$5.71[8]$
\\
$w^{(\gamma,e^+e^-)}_{i\to2p_{1/2}\to1s}$& &$1.67[1]$&$1.22[7]$&$3.71[1]$&$2.04[10]$&$2.61[6]$&$6.79[3]$&$4.70[8]$  
\\
$w^{(\gamma,e^+e^-)}_{i\to2p_{3/2}\to1s}$& &&$1.63[7]$&$5.90[6]$&$4.57[9]$&$2.93[6]$&$2.75[10]$&$1.17[9]$  
\\
$w^{(\gamma,e^+e^-)}_{i\to2s\to1s}$ &&&&$2.10[10]$&$2.10[7]$&$2.05[10]$&$2.11[7]$&$2.27[2]$  
\\
%%%%%%%%%%%%%%%%%%%%%%%%%%%%%%%%%%%%%%%%%%%%%%
\hline
\multicolumn{9}{c}{$Z=36\,,\qquad \dee_{1s}^{(\mu)}=200.7596$}
\\
$\Delta\dee^{(\mu)}_{i\to1s}$& $4.1839$&$4.2119$&$4.3563$&$5.2018$&$5.2093$&$5.2099$&$5.2124$&$5.2536$  
\\
$W_i$& $1.75[17]$&$1.74[17]$&$2.46[14]$&$4.96[16]$&$2.34[16]$&$5.04[16]$&$2.29[16]$&$6.91[14]$  
\\
$w^{(e^+e^-)}_{i\to1s}$& $1.20[14]$&$1.23[14]$&$1.62[11]$&$4.08[13]$&$1.53[11]$&$4.28[13]$&$1.52[11]$&$7.07[10]$  
\\
$w^{(e^+e^-)\txt{C}}_{i\to1s}$&$2.47[13]$&$2.55[13]$&$1.61[11]$&$7.27[12]$&$2.96[10 $&$7.74[12]$&$3.00[10]$&$7.04[10]$
\\
$w^{(\gamma,e^+e^-)}_{i\to2p_{1/2}\to1s}$& &$2.20[5]$&$7.60[10]$&$2.78[5]$&$1.33[13]$&$3.84[9]$&$2.32[7]$&$1.23[11]$  
\\
$w^{(\gamma,e^+e^-)}_{i\to2p_{3/2}\to1s}$& &&$9.59[10]$&$7.79[9]$&$2.67[12]$&$3.85[9]$&$1.60[13]$&$3.35[11]$  
\\
$w^{(\gamma,e^+e^-)}_{i\to2s\to1s}$ &&&&$5.75[12]$&$9.50[9]$&$5.57[12]$&$9.66[9]$&$4.74[5]$  
\\
%%%%%%%%%%%%%%%%%%%%%%%%%%%%%%%%%%%%%%%%%%%%%%
\hline
%\\
\multicolumn{9}{c}{$Z=54\,,\qquad \dee_{1s}^{(\mu)}=195.2742$}
\\
$\Delta\dee^{(\mu)}_{i\to1s}$& $7.3448$&$7.4593$&$8.0501$&$9.6624$&$9.6805$&$9.6941$&$9.6964$&$9.8765$  
\\
$W_i$& $6.47[17]$&$6.54[17]$&$7.08[15]$&$1.69[17]$&$1.21[17]$&$1.79[17]$&$1.15[17]$&$1.44[15]$  
\\
$w^{(e^+e^-)}_{i\to1s}$& $9.60[14]$&$1.01[15]$&$3.04[12]$&$2.41[14]$&$3.75[12]$&$2.71[14]$&$3.66[12]$&$1.15[12]$  
\\
$w^{(e^+e^-)\txt{C}}_{i\to1s}$&$1.42[14]$&$1.52[14]$&$3.01[12]$&$3.05[13]$&$4.72[11]$&$3.55[13]$&$4.84[11]$&$1.13[12]$
\\
$w^{(\gamma,e^+e^-)}_{i\to2p_{1/2}\to1s}$& &$3.22[7]$&$4.64[12]$&$3.15[7]$&$1.48[14]$&$9.54[10]$&$1.34[9]$&$4.54[9]$  
\\
$w^{(\gamma,e^+e^-)}_{i\to2p_{3/2}\to1s}$& &&$6.13[12]$&$1.91[11]$&$2.90[13]$&$9.42[10]$&$1.75[14]$&$3.19[11]$  
\\
$w^{(\gamma,e^+e^-)}_{i\to2s\to1s}$ &&&&$1.81[13]$&$4.06[10]$&$1.76[13]$&$4.26[10]$&$9.72[6]$
\\
%%%%%%%%%%%%%%%%%%%%%%%%%%%%%%%%%%%%%
\hline
%\\
\multicolumn{9}{c}{$Z=92\,,\qquad \dee_{1s}^{(\mu)}=183.183$}
\\
$\Delta\dee^{(\mu)}_{i\to1s}$& $12.160$&$12.607$&$15.314$&$18.579$&$18.392$&$18.692$&$18.521$&$19.561$  
\\
$W_i$& $2.13[18]$&$2.32[18]$&$2.09[17]$&$4.88[17]$&$9.50[17]$&$5.71[17]$&$8.85[17]$&$6.21[16]$  
\\
$w^{(e^+e^-)}_{i\to1s}$& $4.72[15]$&$5.51[15]$&$1.05[14]$&$7.28[14]$&$1.02[14]$&$9.96[14]$&$9.84[13]$&$4.10[13]$  
\\
$w^{(e^+e^-)\txt{C}}_{i\to1s}$&$6.04[14]$&$7.22[14]$&$1.02[14]$&$8.43[13]$&$1.02[14]$&$9.66[12]$&$1.20[14]$&$3.95[13]$
\\
$w^{(\gamma,e^+e^-)}_{i\to2p_{1/2}\to1s}$& &$2.78[9]$&$1.92[14]$&$3.70[9]$&$1.71[15]$&$3.08[12]$&$1.10[11]$&$2.56[13]$  
\\
$w^{(\gamma,e^+e^-)}_{i\to2p_{3/2}\to1s}$& &&$2.91[14]$&$6.09[12]$&$3.24[14]$&$3.03[12]$&$2.00[15]$&$2.71[13]$  
\\
$w^{(\gamma,e^+e^-)}_{i\to2s\to1s}$ &&&&$1.15[14]$&$2.16[11]$&$1.20[14]$&$2.72[11]$&$6.74[8]$ \\
%%%%%%%%%%%%%%%%%%%%%%%%%%%%%%%%%%%%%
\end{tabular}
\end{ruledtabular}
\end{table*}

%%%%%%%%%%%%%%%%%%%%%%%%%%%%%%%%

%/home/olyuan/python/py-muon/muon-fls-05.py
\begin{figure}[h]
\includegraphics[width=20.0pc]{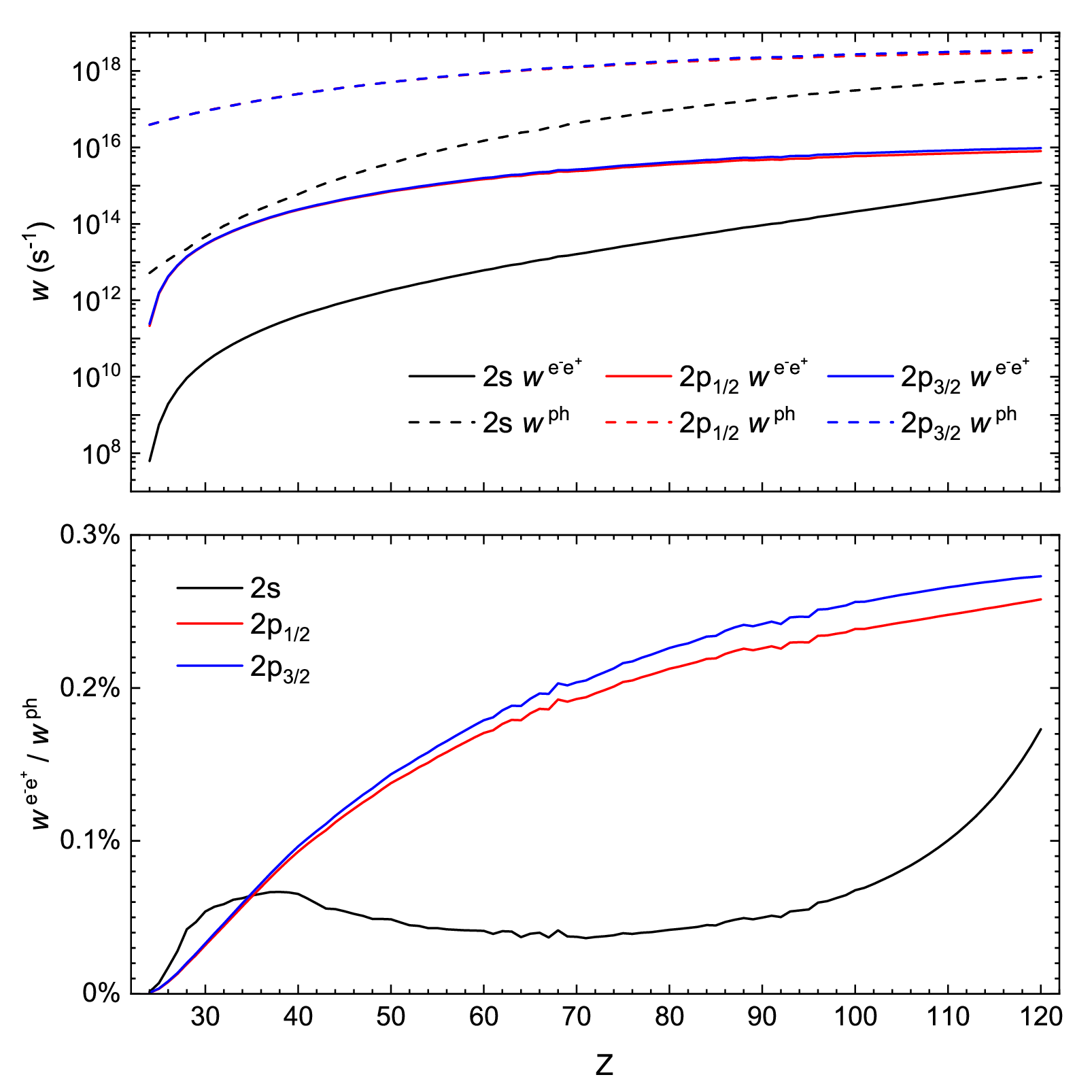}
\caption{
Transition probabilities for excited muon states ($i$): $2p_{1/2}$ (red curves), $2p_{3/2}$ (blue curves) and $2s$ (black curves).
On the upper panel, the radiative transition probabilities ($W_{i}$ in s$^{-1}$, dashed curves) and the nonradiative electron-positron pair production transition probabilities ($w^{(e^{+} e^{-})}_{i\to1s}$ in s$^{-1}$, solid curves) as functions of the nucleus atomic number $Z$.
The bottom panel shows the ratio between the nonradiative and radiative transition probabilities ($w^{(e^{+} e^{-})}_{i\to1s}/W^{}_{i}$).
}
\label{fig-w}
\end{figure}

The transition probabilities for the decay of the $2s$-muon states with electron-positron pair production were considered by Wheeler \cite{wheeler1949RevModPhys.21.133} within the nonrelativistic limit and for point-like nucleus, where the following estimates were presented (our results are given in parentheses) (in s$^{-1}$):
$w^{(e^+e^-, Z=25)}_{2s\to1s}\approx 2[6]$ ($5.54[8]$),
$w^{(e^+e^-, Z=30)}_{2s\to1s}\approx 5[10]$ ($2.46[10]$),
$w^{(e^+e^-, Z=35)}_{2s\to1s}\approx 8[11]$ ($1.25[11]$),
$w^{(e^+e^-, Z=40)}_{2s\to1s}\approx 5[12]$ ($3.87[11]$).
We can see, that the $Z$-dependence of the transition probabilities calculated within the QED differs significantly from the prediction of the nonrelativistic limit \cite{wheeler1949RevModPhys.21.133}.

In Fig.~\ref{fig-w}, we present the radiative transition probabilities for the $2p_{1/2}$, $2p_{3/2}$, and $2s_{1/2}$ states, along with the nonradiative electron-positron pair production transition probabilities, as functions of $Z$. The figure also includes branching ratios for transitions involving electron-positron pair production. We can see that for $Z \geq 36$, the branching ratio for the $2p$ states exceeds that for the $2s$ states. This is attributed to two main factors: the importance of nuclear size corrections for bound muons and the very large contribution of the Breit interaction for $2p$-muon states.

Numerically calculated transition probabilities (in s${}^{-1}$) and transition energies (in $\mx c^2$) for one-muon ions are provided in Table~\ref{tab1}. We analyze the decay of low-lying muon states with $n=2$ and $n=3$ for $Z=24$, $36$, $54$ and $92$. The initial one-muon states ($i$) are indicated in the first row of the table.

The total energies of the $1s$-muon states  ($\dee_{1s}^{(\mu)}$), including the rest mass of muon, are given explicitly for each $Z$ value. The rows marked $\Delta\dee^{(\mu)}_{i\to1s}$ show the energy difference $\Delta\dee^{(\mu)}_{i\to1s} = \dee_{i}^{(\mu)} - \dee_{1s}^{(\mu)}$, where $i$ is the corresponding initial state. This difference represents the energy available for electron-positron pair production. Deviations from the Bohr formula are mainly due to nuclear effects. Incorporating nuclear size corrections significantly alters the order and energy differences of one-muon energy levels, resulting in notably different radiative transition probabilities compared to those for one-electron ions \cite{Knyazeva2022PhysRevA.106.012809}.

The rows $W_i$ represent the total radiative transition probability for transitions from state $i$ to all lower states.

The rows labeled $w^{(e^+e^-)}_{i\to1s}$ present the full QED calculation of the transition probabilities for the nonradiative electron-positron pair production transitions from the initial states $i$ to the $1s$-muon state \Eq{eqn230930n03}. The rows labeled $w^{(e^+e^-)\txt{C}}_{i\to1s}$ show these transition probabilities when only the Coulomb interaction is considered. The difference between the exact interelectron interaction and the Coulomb interaction reflects the contribution of the Breit interaction. We observe that the Breit interaction has a minimal effect on the decay of $2s$ and $3s$ muon states. However, it significantly impacts the decay of muon states with orbital momentum $l\geq1$. Specifically, for $2p$ states, the Breit interaction increases the transition probabilities $w^{(e^+e^-)}_{i\to1s}$ by a factor of $2.8$ for $Z=24$ and by a factor of $7.6$ for $Z=92$. These increases in transition probabilities lead to substantial changes in the patterns of electron-positron pair production in muon ions.

In addition to direct electron-positron pair production, described by \Eq{eqn230930n03}, transitions that involve electron-positron pair production alongside photon emission are also significant for many states. In the dominant cascade channel, these transitions can be represented as follows:
\begin{eqnarray}
\mu_{i}
&\to&\label{eqn240917n01}
\mu_{n} + \gamma
\,\to\,
\mu_{1s} + e^{+}(\dee) + e^{+}(\dee') + \gamma(\omega)
\,,
\end{eqnarray}
where $n$ represents an intermediate state between the initial state $i$ and the final $1s$ state. The transition probabilities for these processes are denoted as $w^{(\gamma,e^+e^-)}_{i \to n \to 1s}$ and are provided in the corresponding rows of Table~\ref{tab1} (for $n=2p_{1/2}$, $2p_{3/2}$, and $2s$). Restricting our analysis to the cascade channel, we calculate these transition probabilities as follows:
\begin{eqnarray}
w^{(\gamma,e^+e^-)}_{i\to n\to1s}
&=&\label{eqn240829n02}
w^{(\gamma)}_{i\to n}\frac{w^{(e^{+}e^{-})}_{n\to1s}}{W_{n}}
\,,
\end{eqnarray}
where $W_{n}$ is the total radiative transition probability for the transition from state $n$. These values are provided in Table~\ref{tab1}. Since the Breit interaction significantly increases the transition probabilities $w^{(e^{+}e^{-})}_{n \to 1s}$ (for states with the orbital momentum $l \geq 1$), it also substantially enhances the $w^{(\gamma,e^+e^-)}_{i \to n \to 1s}$ transition probabilities.

Even though the $2s$ state is metastable, the increase in transition probabilities results in most electron-positron pairs being produced in the $2p \to 1s$ transitions. Table~\ref{tab1} and Fig.~\ref{fig-w} show that for ions with $Z \geq 36$, even if the muon initially occupies the $2s$ state, the production of an electron-positron pair in the $2p \to 1s$ transitions is more frequent than in the $2s \to 1s$ transition. The cascade channel $2s\to2p\to1s$ includes two transitions, $2p_{1/2} \to 1s$ and $2p_{3/2} \to 1s$, whose combined probability ($w^{(\gamma,e^+e^-)}_{2s\to 2p_{1/2}\to1s}+w^{(\gamma,e^+e^-)}_{2s\to 2p_{3/2}\to1s}$) is higher than that of the direct $2s \to 1s$ transition ($w^{(e^+e^-)}_{2s\to1s}$).

Assuming that the excited bound muon states are relatively evenly populated in experiments on muon capture by atoms, we can conclude that most electron-positron pairs are produced during the $2p \to 1s$ transitions for all atomic numbers $Z \geq 24$. Specifically, taking into account the cascade transitions described by \Eq{eqn240917n01}, which involve photon emission and electron-positron pair production, from the $3s$, $3p$, and $3d$ states, we obtain that electron-positron pair production predominantly occurs during the $2p \to 1s$ transitions. The corresponding transition probabilities are shown in Table~\ref{tab1}.

Since the total energy of the electron-positron pair produced in a bound-bound muon transition is determined by the energy conservation law, the technique of simultaneously detecting both the electron and positron while analyzing their combined energy can help identify pairs produced in specific bound-bound muon transitions. The time projection chamber (TPC) technique \cite{DOKE1993113} is particularly useful for this purpose and requires the use of gas targets, such as krypton (${}_{36}$Kr) and xenon (${}_{54}$Xe). Data for these gases are provided in Table~\ref{tab1}. Another possible experimental approach to studying this effect is to examine collisions of muons with thin foils of elements with $Z \geq 24$, or foils of light atoms containing impurities of heavier elements. Determining that both particles come from a single electron-positron pair with a fixed total energy can be achieved using coincidence detection methods \cite{Stiebing2004}.

%%%%%%%%%%%%%%%%%%%%%%%%%%%%%%%%%%%%%%%%%%%%%%%%%%%%%%%

In summary, we have explored a mechanism for matter creation through the decay of muonic ions, where the excess energy is emitted as an electron-positron pair. Our calculations, based on QED and incorporating nuclear corrections, reveal that the Breit interaction plays a crucial role in decays of states with orbital angular momentum $l\ge1$, increasing the corresponding transition probabilities by a factor of 3 to 7. Moreover, the inclusion of the Breit interaction significantly alters the pattern of electron-positron pair production in muonic ion decays. Numerical analysis shows that although the $2s$ state is metastable, the dominant contributions to pair production come from the decay of $2p$ states.

%{\bf
While it is generally assumed that atomic processes are primarily governed by the Coulomb interaction, with the Breit interaction providing only a minor correction, we have identified a class of processes in which the Breit interaction plays a dominant role. This enhances our understanding of fundamental interactions and may have important implications for many-particle physics. 

Bound-bound muon transitions involving electron-positron pair production represent a distinct branch of QED that remains largely unexplored and holds the potential for uncovering significant physical phenomena.
%}

D.M.V. expresses gratitude for the hospitality of the Institute of Modern Physics of CAS during her visit.
The work is supported by the National Key Research and Development Program of China under Grant No. 2022YFA1602501 and the National Natural Science Foundation of China under Grant No. 12011530060.
The work of O.Y.A., K.N.L., D.M.V. in part of the calculation of the radial integrals was supported solely by the Russian Science Foundation under Grant No. 22-12-00043.
O.Y.A. and K.N.L. were supported by the Chinese Academy of Sciences (CAS) Presidents International Fellowship Initiative (PIFI) under Grant Nos. 2018VMB0016 and 2022VMC0002, respectively.

%%%%%%%%%%%%%%%%%%%%%%%%%%%%%%%%%%%%%%%%%%%%%%%%%%%%%%%%%%%%%%%%%%%

%%%%%%%%%%%%%%%%%%%%%%%%%%%%%%%%%%%%%%%%%%%%%%%%%%%%%%%%%%%%%%%%%

%\nocite{*}

%\bibliography{bib231128}% Produces the bibliography via BibTeX.

\end{document}